\documentclass[a4paper, 11pt,notoc]{article}
\pdfoutput=1

\usepackage{jcappub}
\usepackage{graphicx}
\usepackage{booktabs}
\usepackage{verbatim}
\usepackage{caption}
\usepackage{subcaption}

\newcommand{\sigv}{\langle \sigma v_{\rm{rel}} \rangle}

\newcommand{\mDM}{m_{\rm{DM}}}
\newcommand{\mmed}{M_{\rm{med}}}

\newcommand{\gDM}{g_{\rm{DM}}}
\newcommand{\gq}{g_q}
\newcommand{\gSM}{g_q}

\definecolor{cerulean}{RGB}{44,150,207}

\RequirePackage[normalem]{ulem} %DIF PREAMBLE
\RequirePackage{color}\definecolor{RED}{rgb}{1,0,0}\definecolor{BLUE}{rgb}{0,0,1} %DIF PREAMBLE
\providecommand{\DIFadd}[1]{{\protect\color{blue}\uwave{#1}}} %DIF PREAMBLE
\providecommand{\DIFdel}[1]{{\protect\color{red}\sout{#1}}}                      %DIF PREAMBLE
\begin{document}
\title{\begin{boldmath} \huge Recommendations on presenting LHC searches for missing transverse energy \\ signals using simplified $s$-channel models \\ of dark matter \vspace{7mm} \end{boldmath}}

% \begin{comment}
% %%% OLD %%%
% \title{Comparing collider Dark Matter searches with Direct and Indirect Detection Experiments using simplified models 
% }
% \end{comment}

%\author[a]{LHC DM WG}
%\affiliation[a]{ LHC DM WG}
%\emailAdd{oliver.buchmueller@cern.ch}

%%%%%%

\author[1,*]{Antonio~Boveia,}
\affiliation[*]{Editor}
\affiliation[1]{CERN, EP Department, CH-1211 Geneva 23, Switzerland}
\emailAdd{antonio.boveia@cern.ch}

\author[2,*]{Oliver~Buchmueller,}
\affiliation[2]{High Energy Physics Group, Blackett Laboratory, Imperial College, Prince Consort Road, London, SW7 2AZ, United Kingdom}
\emailAdd{oliver.buchmueller@cern.ch}

\author[3]{Giorgio Busoni,} 
\affiliation[3]{ARC Centre of Excellence for Particle Physics at the Terascale, School of Physics, University of Melbourne, 3010, Australia}
%\affiliation[3]{SISSA and INFN, Sezione di Trieste, Italy}

\author[4]{Francesco~D'Eramo,} 
\affiliation[4]{UC, Santa Cruz and UC, Santa Cruz, Inst. Part. Phys., USA}

\author[1,5]{Albert~De~Roeck,}
\affiliation[5]{Antwerp University, B2610 Wilrijk, Belgium}

\author[6]{Andrea~De~Simone,}
\affiliation[6]{SISSA and INFN Sezione di Trieste, via Bonomea 265, I-34136 Trieste, Italy}

\author[7,*]{Caterina~Doglioni,}
\affiliation[7]{Fysiska institutionen, Lunds universitet, Lund, Sweden}
\emailAdd{caterina.doglioni@cern.ch}

\author[3]{Matthew~J.~Dolan,}
%\affiliation[8]{ARC Centre of Excellence for Particle Physics at the Terascale, School of Physics, University of Melbourne, 3010, Australia}

\author[8]{Marie-Helene~Genest,} 
\affiliation[8]{LPSC, Universite Grenoble-Alpes, CNRS/IN2P3, France}

\author[9,*]{Kristian~Hahn,}
\affiliation[9]{Department of Physics and Astronomy, Northwestern University, Evanston, Illinois 60208, USA}
\emailAdd{kristian.hahn@cern.ch}

\author[10,11,*]{Ulrich~Haisch,}
\affiliation[10]{Rudolf Peierls Centre for Theoretical Physics, University of Oxford, Oxford, OX1 3PN, United Kingdom}
\affiliation[11]{CERN, TH Department, CH-1211 Geneva 23, Switzerland}
\emailAdd{ulrich.haisch@physics.ox.ac.uk}

\author[1]{Philip~C.~Harris,} 
%\affiliation[1]{CERN, EP Department, CH-1211 Geneva 23, Switzerland}

\author[12]{Jan~Heisig,}
\affiliation[12]{Institute for Theoretical Particle Physics and Cosmology, RWTH Aachen University, Germany}

\author[13]{Valerio~Ippolito,} 
\affiliation[13]{Laboratory for Particle Physics and Cosmology, Harvard University, USA}

\author[14,*]{Felix~Kahlhoefer,}
\affiliation[14]{DESY, Notkestra\ss e 85, D-22607 Hamburg, Germany}
\emailAdd{felix.kahlhoefer@desy.de}

\author[15]{Valentin~V.~Khoze,} 
\affiliation[15]{Institute of Particle Physics Phenomenology, Durham University, United Kingdom}

\author[16]{Suchita~Kulkarni,} 
\affiliation[16]{Institut f\"ur Hochenergiephysik, \"Osterreichische Akademie der Wissenschaften, Austria}

\author[17]{Greg~Landsberg,} 
\affiliation[17]{Physics Department, Brown University, Providence, Rhode Island 02912, USA}

\author[18]{Steven~Lowette,} 
\affiliation[18]{Vrije Universiteit Brussel IIHE, Belgium}

\author[2]{Sarah~Malik,} 
%\affiliation[nFxf]{Imperial College London, United Kingdom}

\author[11,*]{Michelangelo~Mangano,}
%\affiliation[11]{CERN, EP Department, CH-1211 Geneva 23, Switzerland}
\emailAdd{michelangelo.mangano@cern.ch}

\author[19,*]{Christopher~McCabe,}
\affiliation[19]{GRAPPA Centre of Excellence, University of Amsterdam, Science Park 904, 1098 XH Amsterdam, Netherlands}
\emailAdd{c.mccabe@uva.nl}

%\author[8]{Matthew McCullough,} 
% removed by Uli

\author[20]{Stephen~Mrenna,} 
\affiliation[20]{FNAL, USA}

\author[21]{Priscilla~Pani,} 
\affiliation[21]{Stockholm University, Sweden}

\author[1]{Tristan~du~Pree,} 
%\affiliation[a]{CERN, EP Department, CH-1211 Geneva 23, Switzerland}

\author[11]{Antonio~Riotto,} 
%\affiliation[20]{Department of Theoretical Physics, University of Geneva, Switzerland}

\author[19,22]{David~Salek,} 
\affiliation[22]{Nikhef, Science Park 105, 1098 XG Amsterdam, Netherlands}

\author[14]{Kai~Schmidt-Hoberg,}

\author[23]{William~Shepherd,}
\affiliation[23]{Niels Bohr International Academy, Niels Bohr Institute, University of Copenhagen, Blegdamsvej 17, DK-2100 Copenhagen, Denmark}

\author[24,*]{Tim~M.P.~Tait,}
\affiliation[24]{Department of Physics and Astronomy, University of California, Irvine, California 92697, USA}
\emailAdd{ttait@uci.edu}

\author[25]{Lian-Tao~Wang,} 
\affiliation[25]{Enrico Fermi Institute and Department of Physics and Kavli Institute for Cosmological Physics, University of Chicago, USA}

\author[26]{Steven Worm} 
\affiliation[26]{Particle Physics Department, Rutherford Appleton Laboratory, United Kingdom}

\author[27]{and~Kathryn~Zurek} 
\affiliation[27]{University of California and LBNL, Berkeley, USA}

\hfill CERN-LPCC-2016-001

\abstract{
This document summarises the proposal of the LHC Dark Matter Working Group on how to present LHC results on $s$-channel simplified dark matter models and to compare them to direct (indirect) detection experiments.
}

%%% OLD %%%
% \begin{comment}
% \abstract{
% This document contains a proposal of the LHC Dark Matter group~\cite{LHCDMWG} summarising an approach on how to present limits of simplified dark matter models from collider searches in search planes of Direct Detection experiments defined by the dark matter-nucleon scattering cross section $\sigma_{\rm{SD/SI}}$ and the mass of the dark matter candidate $\mDM$ and self-annihilation cross-sections used by indirect dark matter detection experiments. 
% }
% \end{comment}

%\arxivnumber{1405.0495}

\maketitle
%\flushbottom

%%%%%%%%%%%%%%%%
\section{Introduction}

The interpretation of searches for Dark Matter (DM) (or any other LHC
physics result) requires that one assumes a model leading to the
signal under consideration. This is necessary to compare searches across channels, searches at other center-of-mass energies or at other collider
experiments. The ATLAS and CMS experiments at the LHC coordinated
in 2015 a joint forum to address this issue, in collaboration with
theorists. This ATLAS/CMS DM Forum produced 
a report~\cite{Abercrombie:2015wmb}, providing a first set of concrete simplified DM models used by
ATLAS and CMS to interpret their searches for missing transverse
energy~(MET) signatures.

At the end of the DM forum's activities, a formal LHC Dark Matter WG
(LHCDMWG) was created, to continue the discussion and harmonisation of the way
in which the LHC DM results are interpreted, reported and compared to those of
other experimental approaches.

This document provides the LHCDMWG recommendations on how to present the LHC
search results involving the $s$-channel models considered
in~\cite{Abercrombie:2015wmb} and how to compare these results to
those of direct (DD) and indirect detection (ID) experiments. This
document is the result of the discussions that took place during the
first public meeting of the LHCDMWG~\cite{LHCDMWGWorkshop}, and it is
intended to provide a template for the presentation of the LHC results
at the winter conferences in 2016. It reflects the feedback obtained from
the participants and in subsequent iterations with members of the
experiments and of the theory community and it is based on work
described recently
in~\cite{Buchmueller:2014yoa,Abdallah:2014hon,Malik:2014ggr,Buckley:2014fba,Harris:2014hga,Haisch:2015ioa,Abdallah:2015ter}. For
earlier articles discussing aspects of simplified $s$-channel DM
models, see
also~\cite{Petriello:2008pu,Gershtein:2008bf,Dudas:2009uq,Bai:2010hh,Fox:2011pm,Goodman:2011jq,An:2012va,Frandsen:2012rk,Dreiner:2013vla,Cotta:2013jna,Buchmueller:2013dya,Abdullah:2014lla}.

The relevant details of simplified DM models involving vector,
axial-vector, scalar and pseudo-scalar $s$-channel mediators are first
reviewed in Section~\ref{sec:models}.
Section~\ref{sec:colliderresults} presents a recommendation for the
primary treatment of LHC DM bounds and introduces all of the basic
assumptions entering the approach.
Section~\ref{sec:comparisontonon-colliderresults} describes a
well-defined translation procedure, including all relevant formulas
and corresponding references, that allows for meaningful and fair
comparisons with the limits obtained by DD and ID experiments.

%%%%%%%%%%%%%%%%

%%%%%%%%%%%%%%%%
%\section{Models considered}
\section{Models considered}
\label{sec:models}
The recommendations in this proposal, adopt the model choices made for the early Run-2 LHC searches by the ATLAS/CMS DM Forum~\cite{Abercrombie:2015wmb}. 
In this document we discuss
models which
assume that the DM particle is a Dirac fermion~$\chi$ and that the particle mediating the interaction (the ``mediator") is exchanged in the $s$-channel.\footnote{An orthogonal set of models describe $t$-channel exchange \cite{Chang:2013oia,An:2013xka,Bai:2013iqa,DiFranzo:2013vra}. This class of simplified DM models is left for future iterations and will thus not be discussed in the following.}
After simplifying assumptions,
each model is characterised by four parameters: the DM mass~$\mDM$, the mediator mass~$\mmed$, the universal mediator coupling to quarks~$\gq$  and the mediator coupling to DM~$\gDM$. Mediator couplings to leptons are always set to zero in order to avoid the stringent LHC bounds from 
di-lepton searches.  
In the limit of large $\mmed$, these (and all) models converge to a universal set of operators in an effective field theory (EFT)
\cite{Beltran:2010ww,Goodman:2010yf,Bai:2010hh,Goodman:2010ku,Rajaraman:2011wf,Fox:2011pm}.
In this section, we review the models and give the formulas for the total decay width of the mediators in each case. 

\subsection{Vector and axial-vector models}

The two models with a spin-1 mediator $Z'$, have the following interaction Lagrangians
\begin{align}
\label{eq:AV1} 
\mathcal{L}_{\text{vector}}&=- \gDM Z'_{\mu} \bar{\chi}\gamma^{\mu}\chi -   \gq  \sum_{q=u,d,s,c,b,t} Z'_{\mu} \bar{q}\gamma^{\mu}q \,, \\
\label{eq:AV2} 
\mathcal{L}_{\text{axial-vector}}&=- \gDM Z'_{\mu} \bar{\chi}\gamma^{\mu}\gamma_5\chi - \gq \sum_{q=u,d,s,c,b,t} Z'_{\mu} \bar{q}\gamma^{\mu}\gamma_5q\,.
\end{align}
Note that the universality of the coupling $\gq$ guarantees that the above spin-1 simplified models are minimal flavour violating (MFV)~\cite{D'Ambrosio:2002ex}, which is crucial to avoid the severe existing constraints arising from quark flavour physics. 

The minimal decay width of the mediator is given by the sum of the partial widths for all decays into DM and quarks that are kinematically accessible. For the vector mediator, the partial widths are given by
\begin{align}
\Gamma_{\text{vector}}^{\chi\bar{\chi}} & = \frac{\gDM^2 \hspace{0.25mm} \mmed}{12\pi} 
 \left (1-4 \hspace{0.25mm}  z_{\rm{DM}} \right )^{1/2} \left(1 + 2 \hspace{0.25mm}  z_{\rm{DM}} \right) \, , \\
\Gamma_{\text{vector}}^{q\bar{q}} & = \frac{\gq^2 \hspace{0.25mm}  \mmed}{4\pi} 
 \left ( 1-4 \hspace{0.25mm}  z_q \right )^{1/2}   \left(1 + 2 \hspace{0.25mm}  z_q \right) \, ,
\end{align}
where $z_{{\rm{DM}},q} = m_{\mathrm{DM},q}^2/\mmed^2$ and the two different types of contribution to the width vanish for $\mmed < 2 \hspace{0.25mm}  m_{{\rm{DM}},q}$. The corresponding expressions for the axial-vector mediator are
\begin{align}
\Gamma_{\text{axial-vector}}^{\chi\bar{\chi}} & = \frac{\gDM^2 \, \mmed}{12\pi} 
\left ( 1-4 \hspace{0.25mm} z_{\rm{DM}} \right ) ^{3/2} \,, \\ 
  \Gamma_{\text{axial-vector}}^{q\bar{q}} & =  \frac{\gq^2 \, \mmed}{4\pi} 
\left ( 1-4 \hspace{0.25mm} z_q \right ) ^{3/2} \, .
\end{align}

\subsection{Scalar and pseudo-scalar  models}
\label{sub:smodels} 
The two models with a spin-0 mediator $\phi$ are described by 
\begin{align}
\mathcal{L}_{\text{scalar}}&=- \gDM  \hspace{0.25mm}  \phi \bar{\chi}\chi-  \gq  \hspace{0.5mm}  \frac{\phi}{\sqrt{2}}\sum_{q=u,d,s,c,b,t}  y_q \hspace{0.25mm}  \bar{q}q \,, \label{eq:Scalar} \\
\mathcal{L}_{\text{pseudo-scalar}}&=- i \gDM  \hspace{0.25mm}  \phi \bar{\chi}\gamma_5 \chi - i  \gq \hspace{0.5mm} \frac{\phi}{\sqrt{2}}\sum_{q=u,d,s,c,b,t} y_q \hspace{0.25mm}  \bar{q}\gamma_5 q \label{eq:PseudoScalar}\,,
\end{align}
where $y_q=\sqrt{2} m_q/v$ are the SM quark Yukawa couplings with $v\simeq246$~GeV the Higgs vacuum expectation value. These interactions are again compatible with the MFV hypothesis.

In these models, there is a third contribution to the minimal width of the mediator, which arises from loop-induced decays into gluons. For the scalar mediator, the individual contributions are given by
\begin{align}
 \Gamma^{\chi\bar{\chi}}_{\text{scalar}} & = \frac{\gDM^2 \hspace{0.25mm} \mmed}{8\pi}\left(1 - 4  \hspace{0.25mm}  z_{\rm{DM}}^2\right)^{3/2}\,, \\
 \Gamma^{q\bar{q}}_{\text{scalar}} & =  \frac{3 \hspace{0.25mm}  \gq^2 \hspace{0.25mm}  y_q^2 \, \mmed}{16\pi}\left(1 - 4 \hspace{0.25mm}  z_q^2\right)^{3/2}\, , \\
 \Gamma^{gg}_{\text{scalar}} & = \frac{\alpha_s^2  \hspace{0.25mm}  \gq^2 \hspace{0.25mm}   \mmed^3}{32\pi^3 \hspace{0.25mm}  v^2} \, \big |f_{\rm scalar} (4 \hspace{0.25mm}  z_t)\big |^2 \,, 
\end{align}
while the corresponding expressions in the pseudo-scalar case read 
\begin{align}
 \Gamma^{\chi\bar{\chi}}_{\text{pseudo-scalar}} & =  \frac{\gDM^2 \hspace{0.25mm} \mmed}{8\pi}\left(1 - 4  \hspace{0.25mm}  z_{\rm{DM}}^2\right)^{1/2}\,, \\
 \Gamma^{q\bar{q}}_{\text{pseudo-scalar}} & = \frac{3 \hspace{0.25mm}  \gq^2 \hspace{0.25mm}  y_q^2 \, \mmed}{16\pi}\left(1 - 4 \hspace{0.25mm}  z_q^2\right)^{1/2}  \,, \\
 \Gamma^{gg}_{\text{pseudo-scalar}} & =\frac{\alpha_s^2  \hspace{0.25mm}  \gq^2 \hspace{0.25mm}   \mmed^3}{32\pi^3 \hspace{0.25mm}  v^2} \, \big |f_{\text{pseudo-scalar}} (4 \hspace{0.25mm}  z_t)\big |^2 \,.
\end{align}
Here the form factors take the form 
\begin{align}
 f_{\text{scalar}} (\tau) & = \tau\left[1+(1-\tau)  \hspace{0.25mm} \text{arctan}^2 \left(\frac{1}{\sqrt{\tau-1}}\right) \right] \,, \\
 f_{\text{pseudo-scalar}} (\tau) & = \tau \hspace{0.5mm} \text{arctan}^2 \left(\frac{1}{\sqrt{\tau-1}}\right) \,. \label{eq:fPS}
\end{align}
Note that $ f_{\text{scalar}} (\tau)$ and $ f_{\text{pseudo-scalar}} (\tau)$ are still defined for $\tau < 1$, but in this case the form factors are complex. The tree-level corrections to the total widths of the mediator again do not contribute if $\mmed < 2 \hspace{0.25mm}  m_{{\rm{DM}}, q}$, meaning that the corresponding final state cannot be produced on-shell.  Decays to gluon pairs are only relevant for mediator masses between roughly $200 \, {\rm GeV}$ and $400 \,{\rm GeV}$ and if invisible decays are kinematically forbidden.

\section{Presentation of LHC results}
%\section{Presentation of collider results}
\label{sec:colliderresults}

The simplified DM models  defined in the last section  aim at capturing  accurately  the characteristics of MET production at high-energy colliders. They can be understood as a limit of a more general new-physics scenario, where all but the lightest dark-sector states are assumed to be sufficiently decoupled, so that only the interactions that are relevant at LHC energies are interactions between the mediator and DM as well as the SM quarks. 
Aside from this important caveat, a presentation of collider bounds in the simplified model framework requires no further assumptions, meaning  that LHC searches can be used to set model-independent bounds on the parameter space of a simplified model, and that the constraints arising from different channels --- e.g.~mono-jets and di-jets --- can be directly compared~(see for instance~\cite{Chala:2015ama}).  In this section, we spell out the model choices underlying the LHC limits and the relic density calculations.  
Issues arising in the DD and ID context are discussed in subsequent sections.

\subsection{Mass-mass plane}

\label{mass-massplane}

\begin{figure}[!t]
	\centering
	\includegraphics[width=0.6 \textwidth]{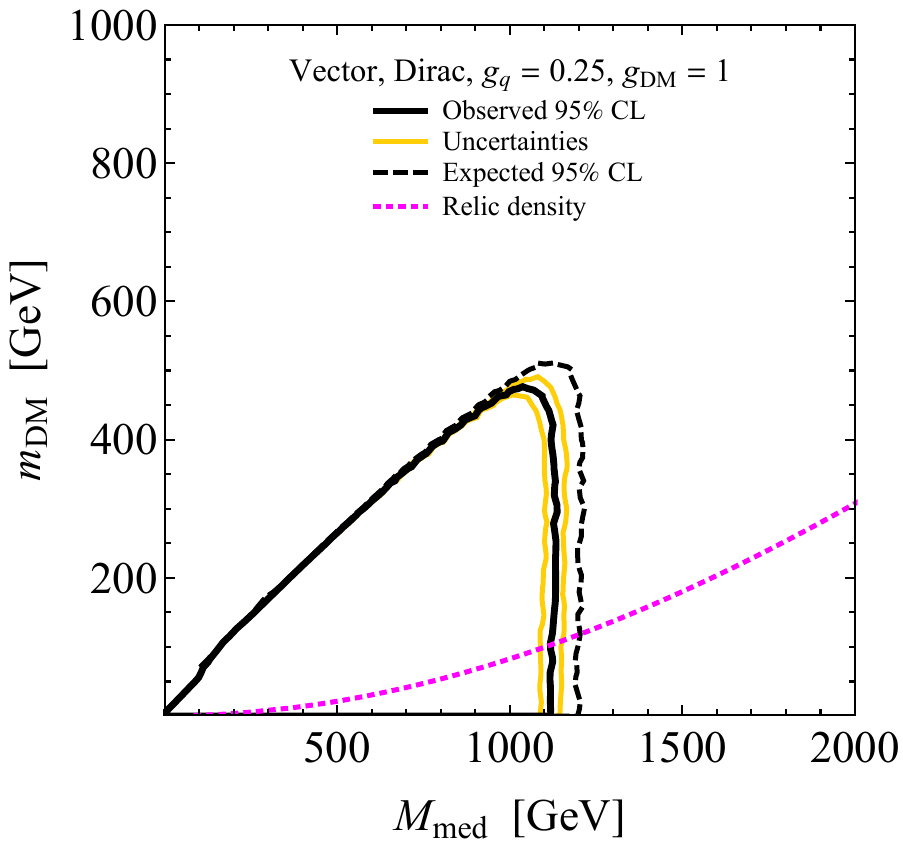}
	\caption{95\% CL exclusion contours in the mass-mass plane for a simplified model with a vector mediator, Dirac DM and couplings $\gq = 0.25$ and $\gDM = 1$. The black solid~(dashed) curve shows the median of the observed (expected) limit, while the yellow curves indicate an example of the uncertainties on the observed bound.  A minimal width is assumed and the excluded parameter space is to the bottom-left of all contours. The dotted magenta curve corresponds to the parameters where the correct DM relic abundance is obtained  from standard thermal freeze-out for the chosen couplings. DM is overproduced to the bottom-right of the curve. The shown  LHC results are intended for illustration only and are not based on real data. }   
	\label{fig:massmass}
\end{figure}

The advocated plots represent only two dimensional slices of the full four dimensional parameter space of the proposed simplified models. To allow for a qualitative understanding of the dependence of the results on the mediator couplings $\gq$ and $\gDM$,  we advocate an auxiliary figure that shows the limit on the ``signal strength'' $\mu$,~i.e.~the ratio of the experimental limit to the predicted signal cross section for fixed masses or fixed coupling scenarios.  We recommend however to clarify that a limit on $\mu$ must not be confused with a bound on a rescaling factor for the couplings and thus in general cannot be used to translate the exclusion limit in the mass-mass plane from one set of couplings to another. The reason is that changing $\gq$ and $\gDM$ typically modifies the total width of the mediator, which can change the kinematic distributions of the signal and thus the exclusion bounds in a non-trivial way. 
Furthermore, for scenarios where the mediator widths varies significantly as the function of the considered parameter (e.g. mass-mass plane), we suggest to add supporting material that illustrates the variation of the width in these parameters.

The primary presentation recommended for LHC results in the simplified model language are plots of the experimental confidence level (CL) limits on the signal cross sections as a function of the two mass parameters $\mDM$ and $\mmed$  for a fixed set of couplings $\gq$ and~$\gDM$. An example of such a ``mass-mass" plot is given in Figure~\ref{fig:massmass}. It shows 95\%~CL exclusion limits (black and yellow curves) for the case of a vector mediator. The limits are derived from a hypothetical LHC mono-jet measurement. The particular choice of axes, with
 $\mmed$ on the $x$-axis and $\mDM$ on the  $y$-axis, follows the convention adopted when interpreting supersymmetry 
 searches at the LHC. The parameter space shown in the mass-mass plots can be divided into three regions: 

\begin{description}

\item[On-shell region:] The on-shell region, $\mmed > 2  \hspace{0.25mm}  \mDM$, is the region where  LHC searches for MET signatures provide the most stringent constraints. The production rate of the mediator decreases with increasing $\mmed$ and so does the signal strength in mono-jet searches. In this region the  experimental limits and the signal cross sections depend in a complex way on all parameters of the simplified model, and it is therefore in general not possible to translate the CL limit obtained for one fixed set of couplings $\gq$ and $\gDM$ to another by a simple rescaling procedure. 
	
 \item[Off-shell region:] In the off-shell region, $\mmed < 2 \hspace{0.25mm} \mDM$, pair-production of DM particles turns off and the constraints from MET searches rapidly lose power. The cross sections become proportional to the combination $\gq^2  \hspace{0.25mm} \gDM^2$ of couplings, so that in principle the LHC exclusions corresponding to different coupling choices can be derived by simple rescalings.  Deviations from this scaling are observed on the interface between on-shell and off-shell regions $\mmed \simeq 2 \hspace{0.25mm} \mDM$~\cite{Jacques:2015zha}. 
Note that for $\mmed < 2 \hspace{0.25mm} \mDM$ an on-shell mediator will always decay back to SM particles, meaning that the off-shell region can be probed by non-MET searches such as di-jets or di-tops. We also note that if the mediator is light and very weakly coupled to the SM quarks, constraints from DD and/or ID on these models may be typically stronger than those from the LHC.

\item[Heavy mediator limit:] The DM EFT limit is approached as the mediator mass $\mmed$ becomes large. In this limit the mono-jet cross sections scale with the fourth inverse power of the effective suppression scale $M_\ast =  \mmed/\sqrt{\gq \hspace{0.25mm} \gDM}$. For perturbative couplings (i.e.~$\sqrt{\gq  \hspace{0.25mm}  \gDM} \ll 4 \pi$), the EFT results apply to mediators with masses in the multi-TeV range. 

\end{description}

As in the template plot, any presentation of  the LHC limits has to clearly state the model assumptions made to obtain the exclusion contours. We thus advocate to explicitly specify on the figure the simplified model, including the mediator and DM type, and the choices of couplings. 
Besides the observed exclusion bound, the median of the expected limit and uncertainties (e.g.~those arising from scale variations or ambiguities related to the choice of parton distribution functions, as well as experimental uncertainties) are useful information that can be added to these plots. All these ingredients have been included in~Figure~\ref{fig:massmass}.

The usefulness of the bound on $\mu$ is thus limited to cases where kinematic distributions are  the same for different realisations of the simplified model. Such a situation is realised for example in the on-shell region if all couplings are sufficiently small, so that the total decay  width of the mediator obeys $\Gamma_{\rm med} \lesssim 0.3 \hspace{0.25mm} \mmed$. Under these circumstances, one can use the narrow-width approximation~(NWA) to  show, for example, that in the case of a spin-1 mediator the mono-jet  cross section $\sigma ( p p \to \chi \bar \chi + j)$ factorises into mediator production $\sigma ( p p \to Z' + j)$ and the  invisible branching ratio ${\rm Br} (Z^\prime \rightarrow \chi\bar{\chi})$.  This factorisation implies that a bound on $\mu$ can be used to infer a limit  on the invisible branching ratio ${\rm Br} (Z^\prime \rightarrow \chi\bar{\chi})$ of the spin-1 mediator relative to the one in the benchmark model without regenerating  the underlying signal Monte Carlo (MC). Since the NWA can be an imperfect approximation even for weak couplings $\gq$ and $\gDM$ (see for instance~\cite{Jacques:2015zha}), we recommend that care is taken if relying on this argument.

If readers would like to reinterpret experimental limits for different coupling choices, it is their responsibility to ensure that kinematic distributions remain unchanged. To make this issue clear, we recommend that captions of plots showing limits on $\mu$  include a statement along the lines of ``Note that the bound on $\mu$ only applies to coupling combinations that yield the same kinematic distributions as the benchmark model.''.

\subsection{Choice of couplings for presentation of results in mass-mass plane}

\label{sub:couplingsandscaling}

At present, we recommend that mono-jet-like searches produce limits for a single choice of couplings. The ATLAS/CMS DM Forum report~\cite{Abercrombie:2015wmb} forms the basis of our recommendations for the simplified models given in Section~\ref{sec:models}. In particular, we advocate the following coupling values to produce the limits on signal strengths:
\begin{description}
\item[Vector mediator:] $\gDM=1$ and $\gq=0.25$.
\item[Axial-vector mediator:] $\gDM=1$ and $\gq=0.25$.
\item[Scalar mediator:] $\gq=1$ and $\gDM=1$.
\item[Pseudo-scalar mediator:] $\gq=1$ and $\gDM=1$.
\end{description}
The quark coupling $\gq$ should be universal in all cases and the width of the mediator should be set to the minimal width, meaning that it is assumed that the mediator has no couplings other than $\gq$ and $\gDM$.\footnote{Using the same value of $\gq$ for all quarks is theoretically well motivated for the vector, scalar and pseudo-scalar mediator. For the axial-vector mediator, it would also be interesting to consider $g_u = g_c = g_t = - g_d = -g_s = -g_b$, which arises naturally if the vector mediator corresponds to the massive gauge boson of a new broken $U(1)^\prime$ and the SM Yukawa couplings are required to be invariant under this additional gauge symmetry. The relative sign between the coupling to up-type 
and down-type quarks is important if interference plays a role and affects the comparison between LHC and  DD results.} The choices above provide for a consistent comparison across collider results within a given simplified model. They ensure that the mediator has $\Gamma_{\rm med}/M_{\rm med} \lesssim 10\%$ and that the theory is far from the strong coupling regime. The choice of $\gq=0.25$ for spin-1 mediators is furthermore motivated by the requirement to avoid di-jet constraints from the LHC and earlier hadron colliders (see e.g.~\cite{Chala:2015ama}). When readers are interested in extrapolating the provided results to other coupling values, it is their responsibility to understand how changing $\gq$ and $\gDM$ will affect the kinematics of the signal and therefore the experimental CL limits. To facilitate such an extrapolation, ATLAS and CMS could provide additional information (e.g.~tables of acceptances, efficiencies, number of events generated, total experimental uncertainty, number of events passing analysis cuts for benchmark signals)
corresponding to the recommended coupling choices as supplementary material, as detailed in Appendix B of~\cite{Abercrombie:2015wmb}. 
As discussed in~\cite{Abercrombie:2015wmb}, the kinematics of vector and axial-vector models is very similar in the case of jet radiation. The same consideration applies for the scalar and pseudo-scalar models in the mono-jet channel, while differences are seen for heavy flavour final states. 

\subsection{Overlaying additional information on LHC results}
\label{sec:overlayingadditionalinformation}

Fixing both $\gq$ and $\gDM$ has the advantage that, in a given model, one can compare the LHC results to relic density calculations or the limits obtained by  DD and ID experiments. Nevertheless, such comparisons typically 
require additional assumptions and should be done carefully. We discuss a few possibilities below. In all cases, we recommend to keep the plots simple, and to specify the assumptions clearly or to produce several variations to indicate the impact that different assumptions have on the final results. 

\subsubsection{Relic density}
\label{sub:relicdensity}

Relic density calculations can be overlaid on the mass-mass plot to indicate where the particles and interactions of a specific simplified model are by themselves sufficient for explaining the observed DM abundance.  For the simplified models recommended by the ATLAS/CMS DM Forum, this curve corresponds to the parameters for which the observed relic abundance is compatible with a single species of DM Dirac fermion and a single mediator that couples to all SM quarks with equal strength. One should not conclude that a simplified model is ruled out for values of model parameters that are inconsistent with the relic density overlay. Rather, one should conclude that additional physics beyond the simplified model was relevant for determining the DM abundance in the early Universe. 

When calculating the relic density, we recommend to include all tree-level processes relevant for the DM annihilation.  In particular, when $\mmed < \mDM$,
annihilation into on-shell mediators are typically active, and are particularly important when $\gDM \gg \gSM$ (e.g.~\cite{Abdullah:2014lla}), for
which cross sections are typically insensitive to $\gSM$, unlike LHC processes.

Numerical tools, such as {\tt micrOMEGAs}~\cite{Belanger:2014vza} and {\tt MadDM}~\cite{Backovic:2015cra}, can be used to calculate the regions of relic overproduction or underproduction for the simplified models recommended by the ATLAS/CMS DM Forum. We provide the results of {\tt MadDM}~calculations for the models described in Section~\ref{sec:models} at~\cite{relic_results}.  These results were obtained using the coupling values specified in Section~\ref{sub:couplingsandscaling}.  The reader should be aware that the axial-vector calculation does not include an explicit constraint from perturbative unitarity (described below).  The provided curves correspond to $\Omega_{\chi}h^{2}=0.12$ (the relic DM density observed by WMAP~\cite{Hinshaw:2012aka} and Planck~\cite{Ade:2015xua}) for the models considered.  Larger mediator masses as well as smaller DM masses (below the curves) correspond to larger values of $\Omega_{\chi}h^{2}$ (and conversely for smaller mediator masses and larger DM masses).

\subsubsection{Perturbativity limits, anomalies and issues with gauge invariance}
\label{sub:perturbativitylimit}

The couplings recommended by the ATLAS/CMS DM Forum have been fixed to values which are perturbative, with the mediator width always sufficiently smaller than the mediator mass. However, it was shown in~\cite{Chala:2015ama, Kahlhoefer:2015bea} that perturbative unitarity is violated in the axial-vector model due to the DM Yukawa coupling becoming non-perturbative, even for perturbative values of $g_q$ and $g_\text{DM}$, if $\mDM$ is significantly larger than $\mmed$.  It was argued that this consideration implies $m_\text{DM}^2 \gDM^2 / (\pi \mmed^2) < 1/2$,
which yields $m_\text{DM} < \sqrt{\pi / 2} \hspace{0.25mm} \mmed$ for the recommended value $\gDM = 1$. It is therefore proposed to indicate the line corresponding to $\mDM = \sqrt{\pi / 2} \hspace{0.25mm} \mmed$ in the mass-mass plot for the axial-vector case in a similar style as for the relic density constraint (i.e.~just a line, no shading).

Another potential problem of the  vector and axial-vector model  is that they are not anomaly free if the $Z^\prime$ boson  couples only to quarks but not to leptons. This  implies that the full theory that ultraviolet completes (\ref{eq:AV1}) and  (\ref{eq:AV2}) must include new fermions to cancel the anomalies. While these fermions can be vector-like with respect to the SM, they will need to be chiral with respect to the new gauge group that gives rise to the~$Z^\prime$. In consequence, the  additional fermions  must have masses of the order of the symmetry-breaking scale, which  is at most a factor of a few above $M_{\rm med}$ \cite{Kahlhoefer:2015bea}.  While the existence of additional fermions will lead to new signatures, the precise impact on LHC phenomenology depends on the specific way the anomalies are cancelled. The resulting model dependence is difficult to quantify and we thus propose to ignore the issue of anomalies until it has been studied in detail by theorists. 

The interactions between the spin-0 mediator and the quarks present in the simplified scalar model are not $SU(2)_L$ invariant. As a result, these interactions will violate perturbative unitarity at high energies in tree-level process like $pp \to W + \phi \, (\phi \to \chi \bar \chi)$. The corresponding amplitudes are however proportional to the squares of the light-quark Yukawa couplings, so that in practice unitarity-violating effects are expected to have a negligible impact on the outcome of MET searches at the LHC. Still in $SU(2)_L$ invariant theories that provide specific realisation of the $s$-channel scalar mediator interactions~(\ref{eq:Scalar}),  like for instance the fermion singlet DM model (see~e.g.~\cite{Abdallah:2015ter}), the resulting LHC phenomenology can be modified by the new fields that are needed to make the full theory gauge invariant. These modifications are again model dependent and lacking detailed theoretical studies, their effect on the  LHC bounds cannot yet be quantified.

\subsubsection{Additional plots}
\label{additionalplot}

Above, we recommend that LHC searches present the mass-mass plot, fixing both $\gq$ and $\gDM$, as the primary result. If desired, additional information on the coupling dependence of the results can be conveyed by producing a related set of limits where one of the mass parameters and one of the couplings has been fixed, and the other mass parameter and coupling are varied. As discussed in the previous section, a correct treatment of varying couplings is one which correctly accounts for the varying acceptance of the search.

\subsubsection{Non-collider DM searches}
\label{sub:non-colliderdmsearches}

Interpreting non-collider results  in the simplified model framework involves additional assumptions, and generally requires detailed knowledge of how the non-collider results were produced.
For example, as discussed above, the relic density predicted by the simplified model varies from point to point on the mass-mass plot, whereas non-collider results are typically presented under the assumption that the density of the DM particle under consideration saturates the cosmological density (i.e.\ that there is just one species of DM). These assumptions may be consistent if there is additional physics (not captured by the simplified model) that affects the relic density calculation but is irrelevant to the LHC signals~(see e.g.~\cite{Gelmini:2010zh}). However, it is also a possibility that the DM particle probed by non-collider experiments constitutes only a certain component of the DM density, so their results would have to be rescaled accordingly (see~for instance~\cite{Chala:2015ama}). Because of the ambiguity of this rescaling, we do not recommend mapping from non-collider results onto the LHC mass-mass plots. The following section addresses the comparison of LHC and non-collider results.

%%%%%%%%%%%%%%%%

%%%%%%%%%%%%%%%%
\section{Comparison to non-collider results}
\label{sec:comparisontonon-colliderresults}

Although we advocate mass-mass plots as the primary 
presentation of  LHC results, it is nevertheless interesting and informative to compare  the LHC limits with the results from other DM searches. To avoid the difficulties associated with reinterpreting the results of non-collider experiments, we recommend translating the LHC results onto the plots of non-collider experiments, rather than the reverse procedure. When performing a translation to the non-collider planes, it is important to bear in mind the different underlying assumptions.  While the DD or ID bounds may be valid for multiple DM models, the LHC limits hold exclusively for the mediator under investigation and for the specific choices of the couplings used in the simplified model.

For a given mediator the translation procedure is well-defined. In this section, we explain all of the ingredients needed for a correct translation into the cross section-mass planes in which  DD and ID experiments present their results. As input, we use  LHC bounds in the mass-mass plane for fixed couplings~$\gSM$ and~$\gDM$ (see Section~\ref{sec:colliderresults}). To compare with  DD experiments, these limits are translated into the  planes of the DM mass $\mDM$ versus the spin-independent (SI) or spin-dependent (SD)  DM-nucleon cross section, $\sigma_{\rm SI}$ or~$\sigma_{\rm SD}$. For a comparison with ID experiments, the limits are instead converted into the plane defined by $\mDM$ and the DM annihilation cross section $\sigv$.

\subsection{ DD experiments}

DD experiments search for the recoil of a nucleus scattering off a DM particle traversing the detector. Since the DM particle is non-relativistic, the dominant interactions between DM and nuclei can be described by two effective parameters, namely the SI and SD DM-nucleon scattering cross sections. DD experiments present their limits as bounds on these  cross section as a function of $\mDM$, where common units for the  cross section are either $\text{cm}^2$ or~pb. The  bounds are presented at 90\%~CL, as opposed to the 95\%~CL limits that are the standard in the collider community. For the sake of comparison, we recommend to present the  LHC limits on the  $\mDM$--$\hspace{0.25mm} \sigma_{\rm{SI/SD}}$ planes at 90\%~CL.

In principle, it is necessary to distinguish between the DM-proton scattering cross sections  $\sigma^p_{\rm{SI}/\rm{SD}}$ and the DM-neutron scattering cross sections  $\sigma^n_{\rm{SI}/\rm{SD}}$. For  SI interactions, however,  DD bounds are always shown under the assumption that $\sigma^p_{\rm{SI}} = \sigma^n_{\rm{SI}}$, which also holds for the simplified models proposed here. For  SD interactions it is common to present separate bounds on $\sigma^p_{\rm{SD}}$ and $\sigma^n_{\rm{SD}}$ and it is possible to compare  LHC results with both.

There are currently a rather large number of  DD experiments that have different target nuclei and use different detection technologies. For  SI interactions, the most sensitive experiments for DM particles heavier than $\mathcal{O}(10 \, \mathrm{GeV})$ are two-phase xenon experiments. There are two large competing collaborations employing this technology, LUX~\cite{Akerib:2015rjg} and XENON1T~\cite{Aprile:2015uzo} (previously XENON100). LUX has published results from its first run and is currently collecting more data to improve its sensitivity. XENON1T will soon begin its first run and aims to have first results in late 2016. For DM particles lighter than $\mathcal{O}(10 \, \mathrm{GeV})$, solid state cryogenic detectors as used by the SuperCDMS~\cite{Agnese:2015nto} and CRESST-II~\cite{Angloher:2015ewa} collaborations are more constraining than xenon experiments as their energy threshold is lower.

As mentioned above, for  SD interactions, separate bounds are published on $\sigma^p_{\rm{SD}}$ and~$\sigma^n_{\rm{SD}}$. This is because DM scatters with the spin of the isotope which is approximately due to an unpaired neutron or unpaired proton. In practice this means that DD experiments have good sensitivity to $\sigma^p_{\rm{SD}}$ or $\sigma^n_{\rm{SD}}$ but not both. The strongest DD limits on~$\sigma^p_{\rm{SD}}$ are from the PICO collaboration~\cite{Amole:2016pye,Amole:2015pla}, while the strongest limits on $\sigma^n_{\rm{SD}}$ are from LUX~\cite{Akerib:2016lao}.\footnote{Note an open source data resource where many DD experiments have uploaded limits is {\tt DMTools}~\cite{DMTools}. These data, however, are not always officially blessed or scrutinised by the experiments and thus should be used with care.}

The simplified models with a vector and scalar mediator lead to a SI interaction, while the axial-vector and  pseudo-scalar mediator induce SD interactions. The  pseudo-scalar interaction has additional velocity-suppression in the non-relativistic limit, which is not present in the other interactions. In practice this means that  pseudo-scalar interactions are only very weakly testable with  DD experiments. For this reason, we will only describe the translation procedure into the  $\mDM$--$\hspace{0.25mm} \sigma_{\rm{SI/SD}}$ plane for 
vector, axial-vector and scalar interactions.

\begin{figure}
         \centering
	\begin{subfigure}{0.45\textwidth}
		\centering
		\includegraphics[width=1\textwidth]{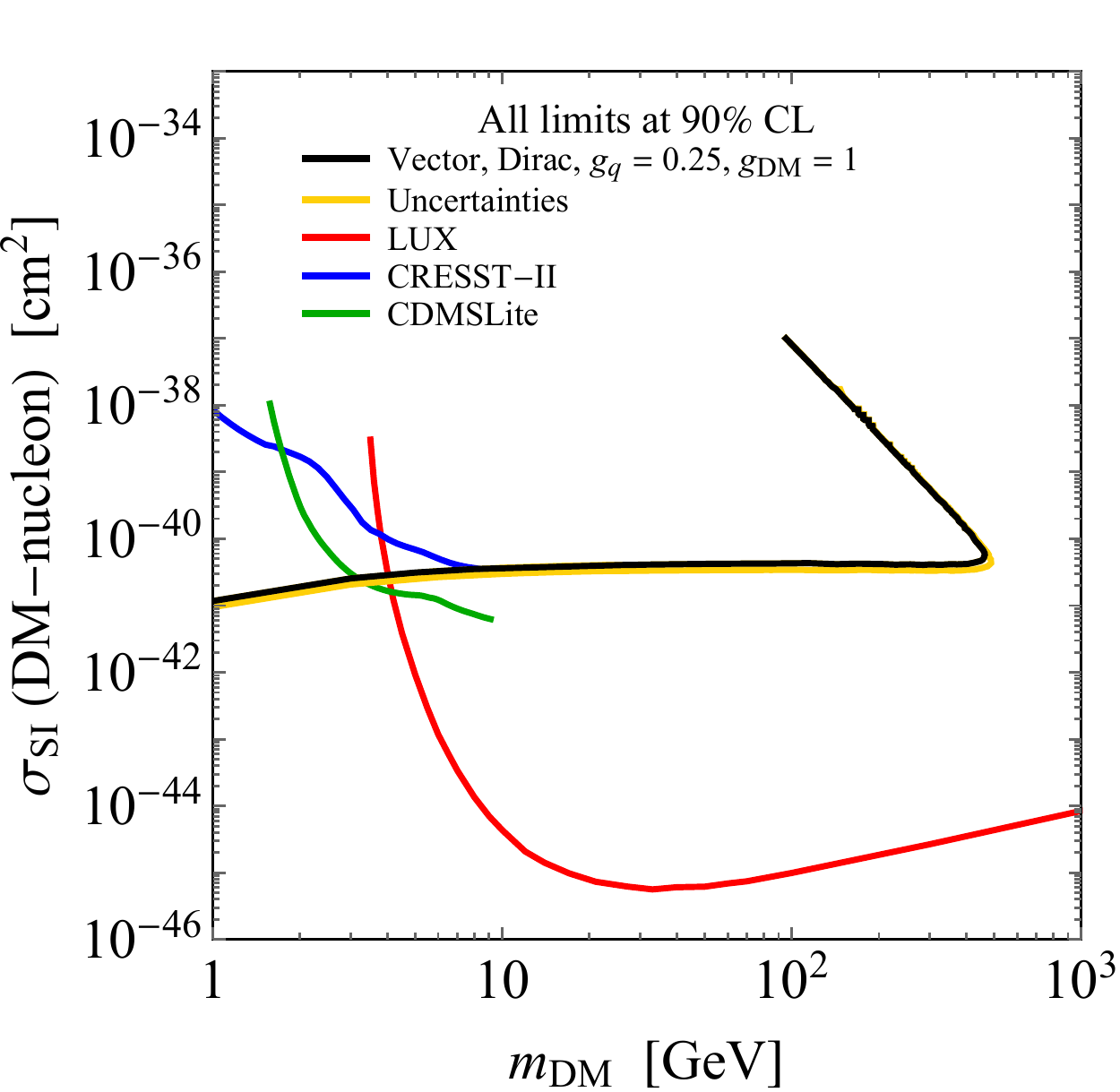}
		\caption{~}
		\label{fig:SI}
	\end{subfigure}
	\qquad 
	\begin{subfigure}{0.45\textwidth}
		\centering
		\includegraphics[width=1\textwidth]{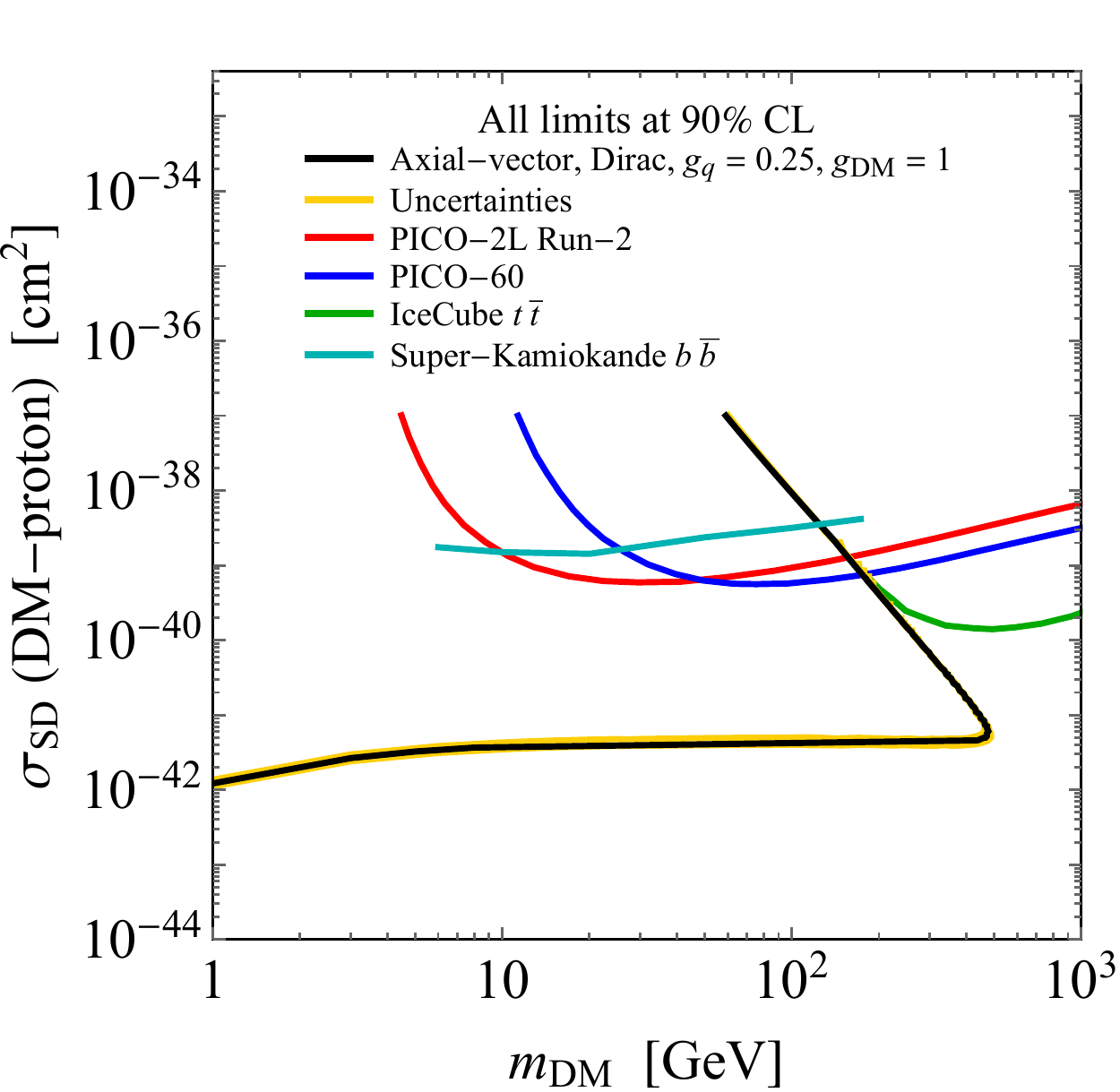}
		\caption{~}
		\label{fig:SD}
	\end{subfigure}
	\caption{		
	A comparison of  LHC results to the  $\mDM$--$\hspace{0.25mm} \sigma_{\rm{SI}}$   (a) and  $\mDM$--$\hspace{0.25mm} \sigma_{\rm{SD}}$ (b) planes. Unlike in the mass-mass plane, the limits are shown at 90\%~CL. The LHC contour in the~SI~(SD) plane is for a vector (axial-vector) mediator, Dirac DM and couplings $\gq=0.25$ and $\gDM=1$. The LHC SI exclusion contour is compared with the LUX, CDMSLite and CRESST-II limits, which are the most constraining in the  shown mass range. The  SD exclusion contour constrains the DM-proton  cross section and is compared with limits from the PICO experiments, the IceCube limit for the $t\bar{t}$ annihilation channel and the Super-Kamiokande limit for the $b\bar{b}$ annihilation channel. The  depicted LHC results are intended for illustration only and are not based on real data.				 
	}   
	\label{fig:SISD}	
\end{figure}

Sections~\ref{sub:spinindependent} and~\ref{sub:spindependent} detail procedures for translating  LHC limits onto to the  $\mDM$--$\hspace{0.25mm} \sigma_{\rm{SI/SD}}$ planes.  Figures~\ref{fig:SI} and~\ref{fig:SD} illustrate the conventions recommended for the presentation of results obtained from these procedures. These plots show the minimum number of DD limits that we recommend to show. Bounds from other experiments may also be included. As in the mass-mass plots, we recommend to explicitly specify details of the mediator and DM type, the choices of couplings and the CL of the exclusion limits. It may also be useful to show theoretical and experimental uncertainties. Generally, the  LHC  searches exclude the on-shell region in the mass-mass plane such that for a fixed value of $\mDM$, the exclusion contour passes through two values of $\mmed$. This means that when translating into the   $\mDM$--$\hspace{0.25mm} \sigma_{\rm{SI/SD}}$ planes, for a fixed value of $\mDM$, the exclusion contour must pass through two values of $\sigma_{\rm{SI/SD}}$. This explains the turnover behaviour   of the  LHC contours observed in Figures~\ref{fig:SI} and~\ref{fig:SD}.

\subsubsection{SI cases: Vector and scalar mediators}
\label{sub:spinindependent}

In general, the  SI DM-nucleon scattering  cross section takes the form
\begin{equation}
\sigma_{\rm{SI}}=\frac{f^2(\gq) \gDM^2 \mu^2_{n\chi}}{\pi \mmed^4}\,,
\end{equation}
where $\mu_{n\chi}= m_n \mDM/(m_n+\mDM)$ is the DM-nucleon reduced mass with $m_n \simeq 0.939 \, \mathrm{GeV}$ the nucleon mass. The mediator-nucleon coupling is $f(\gq)$ and depends on the mediator-quark couplings. For the interactions mediated by vector and scalar particles and for the recommended coupling choices, the difference between the proton and neutron  cross section is negligible. 
 
For the vector mediator, 
\begin{equation}
f(\gq) = 3 \hspace{0.25mm} \gq \,, 
\end{equation} 
and hence
\begin{align}
\label{eq:SI2MSDM}
 \sigma_{\rm{SI}} \simeq 6.9\times 10^{-41}~\mathrm{cm}^2\cdot\left(\frac{\gq \hspace{0.25mm} \gDM}{0.25}\right)^2\left( \frac{1 \, \mathrm{TeV}}{\mmed}\right)^4\left(\frac{\mu_{n\chi}}{1 \, \mathrm{GeV}} \right)^2\,.
\end{align}

For the simplified model with scalar mediator exchange we follow the recommendation of  ATLAS/CMS  DM Forum~\cite{Abercrombie:2015wmb} and assume that the scalar mediator couples to all quarks (like e.g.~the  SM Higgs). In general the formula for $f(\gq)$ is
\begin{equation}
\label{eq:fscalar}
f^{n,p}(\gq) = \frac{m_{n}}{v} \left[\sum_{q=u,d,s}f_q^{n,p} g_q + \frac{2}{27} \hspace{0.5mm} f^{n,p}_{\rm{TG}}\sum_{Q=c,b,t} g_Q \right]\,.
\end{equation}
Here $f^{n,p}_{\rm{TG}}=1-\sum_{q=u,d,s}f_q^{n,p}$. The state-of-the-art values for $f_q^{n,p}$ are from~\cite{Hoferichter:2015dsa} (for $f_u^{n,p}$ and $f_d^{n,p}$) and~\cite{Junnarkar:2013ac} (for $f_s^{n,p}$) and read  $f_u^{n}=0.019$, $f_d^{n}=0.045$ and $f_s^{n}=0.043$. The values for the proton are slightly different, but in practice the difference can be ignored. Substituting these values, we find that numerically
\begin{equation}
 f(\gq)  = 1.16 \cdot 10^{-3} \hspace{0.5mm} \gq \,,
\end{equation}
and therefore the size of a typical  cross section is
\begin{equation}
\sigma_{\rm{SI}} \simeq 6.9\times 10^{-43}~\mathrm{cm}^2\cdot\left(\frac{\gq \hspace{0.25mm} \gDM}{1}\right)^2\left( \frac{125 \, \mathrm{GeV}}{\mmed}\right)^4\left(\frac{\mu_{n\chi}}{1 \, \mathrm{GeV}} \right)^2\,.
\end{equation}

\subsubsection{SD case: Axial-vector mediator}
\label{sub:spindependent}

For the axial-vector mediator, the scattering is  SD and the corresponding  cross section can be written as
\begin{equation}
\sigma_{\rm{SD}}=\frac{3 \hspace{0.25mm} f^2(\gq) \gDM^2 \mu^2_{n\chi}}{\pi \mmed^4}\,.
\end{equation}
In general $f^{p,n}(\gq)$ differs for protons and neutrons and is given by
\begin{equation}
 f^{p,n}(\gq) = \Delta^{(p,n)}_u \, g_u + \Delta^{(p,n)}_d \, g_d + \Delta^{(p,n)}_s \, g_s \,,
\end{equation}
where $\Delta^{(p)}_u= \Delta^{(n)}_d = 0.84$, $\Delta^{(p)}_d = \Delta^{(n)}_u = -0.43$ and $\Delta_s=-0.09$ are the values recommended by  the Particle Data Group~\cite{Agashe:2014kda}. Other values are also used in the literature (see~e.g.~\cite{Ellis:2008hf}) and differ by up to~$\mathcal{O}(5\%)$.

Under the assumption that the coupling~$g_q$ is equal for all quarks, one finds 
\begin{equation}
f(\gq) = 0.32 \hspace{0.25mm} \gq \,,
\end{equation} 
and thus
\begin{align}
\label{eq:SD2MSDM}
 \sigma^{\rm{SD}} \simeq 2.4\times 10^{-42}~\mathrm{cm}^2\cdot\left(\frac{\gq \hspace{0.25mm} \gDM}{0.25}\right)^2\left( \frac{1 \, \mathrm{TeV}}{M_{\rm{med}}}\right)^4\left(\frac{\mu_{n\chi}}{1 \, \mathrm{GeV}} \right)^2\,.
\end{align}
We emphasise that the same result is obtained both for the  SD DM-proton scattering  cross section $\sigma^p_{\rm{SD}}$ and the  SD DM-neutron scattering  cross section $\sigma^n_{\rm{SD}}$. Using \eqref{eq:SD2MSDM} it is therefore possible to map collider results on both parameter planes conventionally shown by  DD  experiments. Should only one plot be required, we recommend comparing  the LHC results to  the DD  bounds on $\sigma^p_{\rm{SD}}$, which is typically more difficult to constrain.

In the future, it is desirable to consider not only the case $g_u = g_d = g_s$, but also the case $g_u = -g_d = -g_s$, which is well-motivated from embedding the simplified model in the SM gauge group and can be included without much additional effort. For $g_u = -g_d = -g_s$ one obtains approximately $f^p(\gq) = 1.36 \, g_u$ and $f^n(\gq) = - 1.18 \, g_u$, i.e.\ the DM-neutron  cross section is slightly smaller than the DM-proton  cross section.\footnote{LHC searches are only sensitive to the relative sign between $g_u$ and $g_d$ if both types of quarks are present in a single process (e.g.~$u \bar{d} \rightarrow u \bar{d} + \chi \bar{\chi}$ or $u \bar{u} \rightarrow d \bar{d} + \chi \bar{\chi}$). Such processes give a  subleading effect in mono-jet searches and are presently not included in the signal computation. As a result, the signal prediction for mono-jets turns out to be independent of the relative sign between the individual quark couplings~\cite{Haisch:2016usn}.}

\subsubsection{Neutrino observatories: IceCube and Super-Kamiokande}

The IceCube~\cite{Aartsen:2016exj} and Super-Kamiokande~\cite{Choi:2015ara} neutrino observatories are also able to constrain the~SI and SD cross sections. When DM particles elastically scatter with elements in the Sun, they can lose enough energy to become gravitationally bound. Self-annihilation of the DM particles produces neutrinos (either directly or in showering) that can be searched for in a neutrino observatory. When the DM capture and annihilation rates are in equilibrium, the neutrino flux depends only on the initial capture rate, which is determined by the SI or SD cross section~\cite{Silk:1985ax}.  

The IceCube and Super-Kamiokande limits on~$\sigma^p_{\rm{SD}}$ are of particular interest as they can be stronger than the corresponding bounds from DD experiments.  The former bounds are however more model dependent, since they depend on the particular DM annihilation channel.  For annihilation only into light quarks, the limits are weaker than DD experiments. For $m_b< m_\text{DM} < m_t$, on the other hand, the dominant annihilation channel of the axial-vector model is to $b \bar{b}$ and Super-Kamiokande sets more stringent constraints than DD experiments for $\mDM<10 \, \rm{GeV}$. For $m_\text{DM} > m_t$, the dominant annihilation channel is to~$t \bar{t}$ and the resulting constraints from IceCube are stronger than DD experiments. Both the Super-Kamiokande and IceCube limits can be shown together with other bounds on the SD DM-proton scattering cross section.  

While strong bounds are obtained for annihilation into bosons or leptons, these couplings are not present in the simplified models considered here. Therefore, we do not recommend showing the IceCube or Super-Kamiokande limits for annihilation into bosons or leptons. Note also that the IceCube bounds may be further modified if the DM particles can directly annihilate into the mediator (see the discussion in~\cite{Heisig:2015ira}). For $\mDM\lesssim4~\mathrm{GeV}$, the effects of DM evaporation from the Sun are important, so placing
limits on~$\sigma^p_{\rm{SD}}$ and~$\sigma_{\rm{SI}}$ from neutrinos coming from the Sun becomes very difficult in this low-mass regime 
(see~e.g.~\cite{Busoni:2013kaa}).

\subsection{ID experiments}
\label{sec:indirect}

For a  pseudo-scalar mediator, the rate at DD experiments is suppressed by additional velocity-dependent terms entering the  cross section. As a result,  DD experiments have very little sensitivity for this scenario and it is not worthwhile to compare  LHC results to the usual bounds on SI and SD  cross sections. Instead,  LHC bounds  can be compared against the limits from  ID experiments. 
For example, Fermi-LAT places 95\%~CL constraints on the self-annihilation  cross section from observations of dwarf spheroidal galaxies~\cite{Ackermann:2015zua}.\footnote{The galactic center is also potentially a promising DM target.  Current observations show an excess of gamma rays which are roughly consistent with a DM signal, but cannot be conclusively identified as such due to poorly understood astrophysical
backgrounds~\cite{TheFermi-LAT:2015kwa}.  The regions of simplified models capable of reproducing this excess are currently regions of particular interest
for collider and direct searches.} Limits are set on the  cross section $\sigv$ to annihilate to a single  particle-anti-particle final state. 

There are a number of subtleties when dealing with these limits. Firstly, all of the  bounds shown in~\cite{Ackermann:2015zua} are for a Majorana fermion. ID annihilation cross section limits for a Dirac fermion are larger by a factor of two and therefore need to be rescaled before they can be compared to the Dirac DM simplified model considered here. Secondly, the limits are for single  particle-anti-particle final states while models typically include more than one final state. For the  pseudo-scalar model, for example, DM annihilates to all quarks with branching ratios approximately proportional to $m_q^2$.
In practice, however, the gamma-ray flux that is observed from annihilating to different quarks (or gluons) is small~\cite{Cirelli:2010xx}. The~Fermi-LAT limits~\cite{Ackermann:2015zua} also demonstrate that there is a negligible difference between the limits on $\sigv$ in $u \bar{u}$ and $b \bar{b}$ final states. We therefore suggest to only show the bound on $u\bar{u}$ from Fermi-LAT in comparison with the calculated bound on the total annihilation cross section, as representative of the limits to final states involving linear combinations of different quarks or gluons. 

\begin{figure}[t!]
	\centering
	\includegraphics[width=0.6\textwidth]{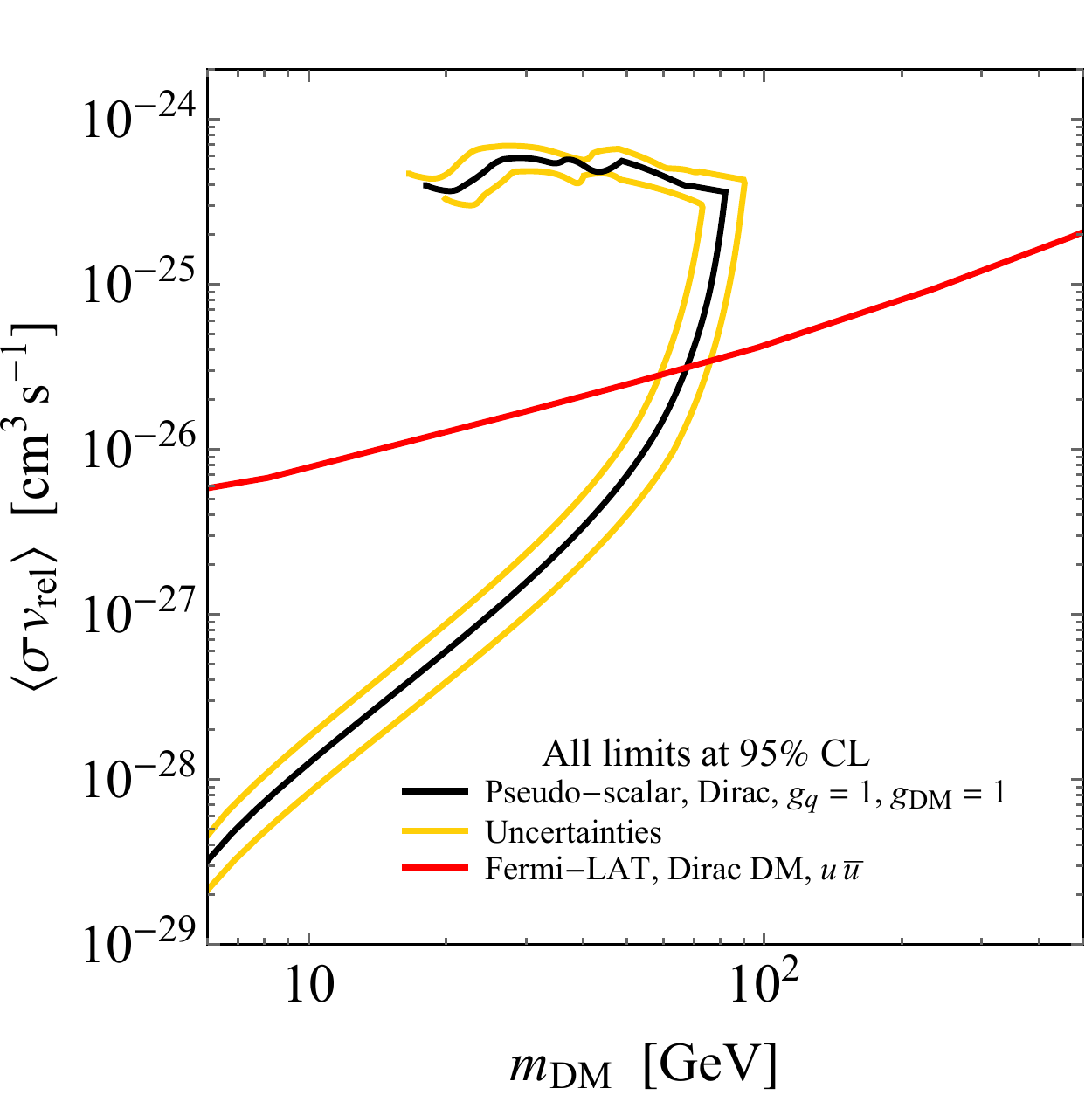}
	\caption{A comparison of the  LHC result to the Fermi-LAT limit in the $\mDM$--$\sigv$  plane. Both limits are at 95\%~CL. The Fermi-LAT limit is for Dirac DM and assumes that the only annihilation channel is to $u \bar{u}$ quarks. The Fermi-LAT limits to other  quark-anti-quark annihilation channels will be similar. The LHC exclusion contour is for a  pseudo-scalar mediator, Dirac DM and couplings $\gq=1$ and $\gDM=1$.  The shown  LHC results are intended for illustration only and are not based on real data.}
	\label{fig:ID}
\end{figure}

The annihilation  cross section into a $q \bar{q}$ final state is (see~e.g.~\cite{Buchmueller:2015eea} for a recent example)  
\begin{equation} \label{eq:qqann}
 \langle \sigma v_{\rm{rel}}\rangle_q =\frac{3 \hspace{0.25mm} m_q^2}{2 \pi v^2} \hspace{0.25mm} \frac{\gq^2  \hspace{0.25mm}  \gDM^2  \hspace{0.25mm}  \mDM^2}{(\mmed^2-4 \mDM^2)^2+\mmed^2 \Gamma_{\rm med}^2} \, \sqrt{1-\frac{m_q^2}{\mDM^2}}\,,
\end{equation}
where  $\Gamma_{\rm med}$ is the total width of the mediator (see  Section~\ref{sub:smodels}). Similarly, the annihilation  cross section into a pair of gluons is given by
\begin{equation} \label{eq:ggann}
\langle \sigma v_{\rm{rel}}\rangle_g =\frac{\alpha_s^2}{2 \hspace{0.25mm}  \pi^3 v^2}\hspace{0.25mm} \frac{\gq^2 \hspace{0.25mm} \gDM^2}{(\mmed^2-4 \mDM^2)^2+\mmed^2 \Gamma_{\rm med}^2} \, \left|\sum_q m_q^2 \hspace{0.5mm} f_\text{pseudo-scalar}\left(\frac{m_q^2}{m_\chi^2}\right)\right|^2\,,
\end{equation}
where $f_\text{pseudo-scalar} (\tau)$ 
has been defined in 
(\ref{eq:fPS}) and $\alpha_s$ is the strong coupling constant, which we recommend to evaluate at the scale $\mu = 2 \mDM$. The  total cross section is then given by  the sum of the quark and gluon channels (\ref{eq:qqann}) and (\ref{eq:ggann}) as well as any annihilation channels into on-shell
mediators which are kinematically allowed and are not suppressed by the small relative velocities of DM in the galactic halo.

Figure~\ref{fig:ID} depicts the translation of  LHC bounds   for a pseudo-scalar mediator to the  $\mDM$--$\sigv$ plane. As with the other plots, we recommend to specify explicitly details including the mediator and DM type, the choices of couplings and the CL of the exclusion limits. It is also important to emphasise that the  ID limit is for Dirac DM instead of Majorana DM as assumed in the Fermi-LAT publication.  Since the  LHC exclusion contour in the mass-mass plane passes through two values of $\mmed$,  the LHC limit shows a similar turnover behaviour in the  $\mDM$--$\sigv$ plane.  In Figure~\ref{fig:ID} we have depicted both branches of the exclusion contour that are obtained for fixed DM mass $\mDM$. 
It may also be desirable to show the values of~$\sigv$ in Figure~\ref{fig:ID} that produce the observed relic density.  A standard reference providing the values of~$\sigv$ as a function of~$\mDM$ is~\cite{Steigman:2012nb}. We reemphasise the point made in~\cite{Steigman:2012nb} that their displayed values of~$\sigv$ should be multiplied by a factor of two for Dirac DM.

To conclude this section, we emphasise that 
translating DD or ID searches into bounds on the DM-nucleon scattering  cross section  or the DM self-annihilation cross section plane always require an assumption on the density of DM particles. In particular, it is always assumed that the particle under consideration constitutes all of the DM in the Universe. If $\chi$ is only one out of several DM sub-components, bounds from  DD and ID experiments would become weaker, while  the LHC bounds  remain unchanged.

%%%%%%%%%%%%%%%%

\section{Acknowledgement} 
This document is part of a project that has received funding from the European Research Council (ERC) under the European UnionÕs Horizon 2020 research and innovation programme (grant agreement No 679305). The work of OB is supported in part by the London Centre for Terauniverse Studies (LCTS), using funding from the European Research Council via the Advanced Investigator Grant 267352. UH acknowledges the hospitality and support of the CERN theory division.  SK is supported by the New Frontiers program of the Austrian Academy of Sciences.  The work of CM is part of the research programme of the Foundation for Fundamental Research on Matter (FOM), which is part of the Netherlands Organisation for Scientific Research (NWO). TMPT acknowledges support from NSF grant PHY-1316792 and UCI through a Chancellor's Fellowship.

%%%Other possible headings, not yet developed
%\subsection{Reinterpretation of collider result}
%\label{reinterpretationofcolliderresult}
%This section could contain references to the collaboration policies for sharing results so that they are easy to reinterpret

\bibliography{ref}
\bibliographystyle{JHEP}

\end{document}